\documentclass[aps,prd,12point,twocolumn,nofootinbib,showpacs,superscriptaddress]{revtex4-2}
\def\theequation{\arabic{section}.\arabic{equation}}
\usepackage{psfrag}
\usepackage{subfigure}
\usepackage{color}
\usepackage{mathrsfs}
\usepackage{graphicx}
\usepackage{amssymb, bm}
\usepackage{amsmath, amsthm}
\usepackage{epstopdf}
\usepackage{hyperref}
\usepackage{enumerate}
\usepackage{longtable}
\usepackage{amsmath}
\usepackage[normalem]{ulem}

\newcommand{\be}{\begin{equation}}
\newcommand{\ee}{\end{equation}}
\definecolor{pinegreen}{rgb}{0.0, 0.47, 0.44}

\begin{document}
\def\theequation{\arabic{section}.\arabic{equation}} 

\title{Tolman-Ehrenfest's criterion of thermal equilibrium extended to  
conformally static spacetimes}

\author{Valerio Faraoni}
\email[]{vfaraoni@ubishops.ca}
%\thanks{}
\affiliation{Department of Physics \& Astronomy, Bishop's University, 
2600 College Street, Sherbrooke, Qu\'ebec, Canada J1M~1Z7}

\author{Robert Vanderwee}
\email[]{rvanderwee20@ubishops.ca}
%\thanks{}
\affiliation{Department of Physics \& Astronomy, Bishop's University, 
2600 College Street, Sherbrooke, Qu\'ebec, Canada J1M~1Z7}

%\date{\today}
\begin{abstract} 

With insight from examples and physical arguments, the Tolman-Ehrenfest 
criterion of thermal equilibrium for test fluids in static spacetimes is 
extended to local thermal equilibrium in conformally static geometries. 
The temperature of the conformally rescaled fluid scales with the inverse 
of the conformal factor, reproducing the evolution of the cosmic microwave 
background in Friedmann universes, the Hawking temperature of the   
Sultana-Dyer cosmological black hole, and a heuristic argument by Dicke.

\end{abstract}

%\pacs{04.50.+h, 04.90.+e}
%alternative theories of gravity
%Other topics in GR and gravitation

%\keywords{thermodynamics of modified gravity, stealth scalar fields, 
%scalar-tensor gravity.}

\maketitle
% If in two-column mode, this environment will change to single-column
% format so that long equations can be displayed. Use
%\begin{widetext}
%\end{widetext}

\section{Introduction}
\label{sec:1}
\setcounter{equation}{0}

Thermal physics in relativity and in curved spacetime is more intriguing, 
and notoriously more difficult, than in non-relativistic situations and 
several results necessarily have limited validity. An example is the 
Tolman-Ehrenfest criterion for the thermal equilibrium of a fluid in a 
static spacetime \cite{Tolman28,Tolman:1930zza,Tolman:1930ona}. In a 
coordinate system adapted to the time symmetry, in which the line element 
reads\footnote{We follow the notation of Ref.~\cite{Wald:1984rg}.}
\be
 ds^2 =g_{00}\left( x^k \right) dt^2 + g_{ij}\left( x^k \right) dx^i dx^j 
\quad\quad  (i,j,k=1,2,3), 
\ee 
the 
temperature ${\cal T}$ of a {\em test} fluid at rest with respect to the 
static observers ({\em i.e.}, those with four-velocity parallel to 
the timelike Killing vector $k^a$) obeys 
\cite{Tolman28,Tolman:1930zza,Tolman:1930ona} \be 
{\cal T} \sqrt{-g_{00}}= {\cal T}_0 \,,\label{Tolman} \ee where ${\cal 
T}_0$ is a constant. Equation~(\ref{Tolman}) is referred to as the 
Tolman-Ehrenfest criterion of thermal equilibrium. It expresses the fact 
that, since heat is mass-energy, it will sink in a gravitational field and 
regions of stronger gravity will be hotter. As a result, a fluid at rest 
in a static gravitational field and in thermal equilibrium has a 
non-vanishing temperature gradient, a counterintuitive result. Klein 
formulated the analogous condition for the equilibrium of particles with 
respect to diffusion in a static spacetime by replacing temperature ${\cal 
T}$ with chemical potential $\mu$ \cite{Klein49}. {\em Caveats} on the 
standard presentations of the Tolman-Ehrenfest criterion have been 
discussed exhaustively in the recent works 
\cite{Santiago:2018kds,Santiago:2018jeu,Santiago:2019aem}, in particular 
the generalization of this law to stationary (but non-static) geometries. 
The criterion has inspired also a connection between gravitational ﬁelds 
and thermal transport in materials: thermal transport, understood as the 
linear response of a material to a temperature gradient, was mimicked by 
Luttinger as a counter-balancing weak gravitational field restoring 
thermal equilibrium in the presence of this gradient \cite{Luttinger:1964zz}. 
The Tolman-Ehrenfest criterion~(\ref{Tolman}) is applied to neutron 
stars \cite{Laskos-Patkos:2022lgy,Kim:2022qlc,Li:2022url}; 
equilibrium with respect to simultaneous heat 
conduction and particle diffusion has been discussed in 
\cite{Lima:2019brf,Kim:2021kou}, together with the corresponding criterion 
in Weyl-integrable geometries \cite{Lima:2021ccv}.

The Tolman-Ehrenfest criterion can be derived from Eckart's generalization 
of the Fourier law for heat conduction, a constitutive relation assumed in 
Eckart's first-order thermodynamics of dissipative fluids 
\cite{Eckart:1940te}. An imperfect fluid with four-velocity $u^a$ is 
described by the energy-momentum tensor
\be
T_{ab} = {\rho}u_a u_b + Ph_{ab} + \pi_{ab} + q_a u_b + q_b u_a 
\,,\label{imperfectfluid}
\ee
where $\rho$ is the energy density, $P$ 
is the isotropic pressure, 
$\pi_{ab}$ is the anisotropic stress tensor,  $q^a $ is the heat flux 
density, and $h_{ab} \equiv g_{ab} + u_a u_b $
is the Riemannian metric on the 3-space orthogonal to $u^a$. $\pi^{ab}$ 
and $q^a$ are purely spatial with respect to 
$u^a$ and $\pi^{ab}$ is trace-free: 
\be
\pi_{ab}u^a  =
\pi_{ab}u^b  =
q^a u_a=0 \,, \quad \quad {\pi^a}_a=0 \,.
\ee
Eckart's theory assumes the three constitutive relations for this fluid 
\cite{Eckart:1940te} 
\begin{eqnarray}
q_a &=& -\mathcal{K} h_{ab} \left( \nabla^b {\cal T} + 
{\cal T} \dot{u}^b  \right) \,, \label{Eckart1}\\
&&\nonumber\\
P &=& P_\mathrm{non-viscous} + P_\mathrm{viscous} \nonumber\\
&&\nonumber\\
&=& P_\mathrm{non-viscous} -\zeta \,\Theta \,,\label{Eckart2}\\
&&\nonumber\\
\pi_{ab} &=& -2\eta \, \sigma_{ab} \,,\label{Eckart3}
\end{eqnarray}
where ${\cal T}$ is the temperature, ${\cal K}$ is the thermal 
conductivity, $\Theta = \nabla_c u^c $ is the expansion scalar, the shear 
tensor $\sigma_{ab} $ is the symmetric, trace-free part of $ {h_a}^c 
{h_b}^d \nabla_d u_c $ \cite{EMMacC}, while $\zeta$ and $\eta$ are the 
bulk and shear viscosity coefficients, respectively. $\dot{u}^c \equiv a^c 
\equiv 
u^b \nabla_b u^c $ is the fluid's four-acceleration, which contributes an 
inertial term to the heat flux~(\ref{Eckart1}) \cite{Eckart:1940te}.

The derivation of the 
Tolman-Ehrenfest criterion from Eq.~(\ref{Eckart1}), which generalizes the 
usual non-relativistic Fourier law of heat conduction, appears in 
\cite[Exercise 22.7, p.~567]{MTW} and, more 
recently, in Ref.~\cite{Santiago:2019aem}. For the reader's convenience, 
we reproduce this derivation in Appendix~\ref{sec:AppendixA}.

We define {\it thermal equilibrium} in a static spacetime (and, later, 
local 
thermal equilibrium in time-dependent ones) as the absence of heat fluxes, 
$q^a=0$. It is clear that, if a fluid is in thermal equilibrium in a 
certain frame, 
any observer moving relatively to it will detect a heat flux (which lies 
at the origin of some of the subtleties in generalizing Eq.~(\ref{Tolman})  
to stationary geometries 
\cite{Santiago:2018kds,Santiago:2019aem,Santiago:2018jeu}). To 
make this observation quantitative, consider a perfect fluid seen from a 
non-comoving frame, in which it appears ``tilted''. Denote (momentarily)  
with a star 
quantities associated with the comoving frame, for example $u^{*a}$ 
is the fluid four-velocity. The stress-energy tensor $T_{ab}$ of the 
perfect fluid (an observer-independent object) can be decomposed according 
to 
this frame as
\be
    T_{ab} = \rho^{*}u^{*}_a u^{*}_b + P^{*} \, h^{*}_{ab} \,,
\ee
where $h^*_{ab} \equiv g_{ab} + u^*_a u^*_b $ is the 
Riemannian three-metric on the 3-space seen by the observers $u^*_a$ 
comoving with the fluid. 

The frame of an observer moving with respect to this fluid (in which the 
fluid appears to be moving) is characterized by a different four-velocity 
$u^{a}$ related to $u^{*a}$ by \cite{EMMacC}
\be
    u^{*a} = \gamma \left( u^{a} + v^{a} \right) \,,
\ee
where $v^a$ is a purely spatial vector according to $u^*_a$, $v_a u^{*a} 
=  0$,  with $ 0 \leq v^2 \equiv  v_a v^a < 1$ and
\be
\gamma =\frac{1}{ \sqrt{1-v^2}} 
\ee
is the corresponding Lorentz factor. The fluid stress-energy 
tensor 
can be decomposed according to the observers\footnote{There is only one 
stress-energy tensor $T_{ab}$ but it can be decomposed in infinitely many 
ways according to the possible timelike observers $u^a$.} $u^a$ as 
\be
 T_{ab} = {\rho} \, u_{a}u_{b} + Ph_{ab} + \pi_{ab} + q_a u_b + 
q_b u_a \,,
\ee
where $ h_{ab} \equiv u_a u_b  + g_{ab}$ and \cite{EMMacC}
\begin{eqnarray}
\rho &=& \rho^{*} + \gamma^2 v^2 \left( \rho^{*} + P^{*} \right) = 
\gamma^2 \left( \rho^{*} + v^2 P^{*} \right) \,, \\
&& \nonumber\\
P &=& P^{*} +  \frac{\gamma^2 v^2}{3} \left( \rho^{*} + P^{*} \right) \,, 
\\
&& \nonumber\\
q^{a} &=& \left( 1 + \gamma^2v^2 \right) \left( \rho^{*} + P^{*} 
\right) v^{a} 
= \gamma^2 \left( \rho^{*} + P^{*} \right) v^a \,, \nonumber\\
&&\label{ABC}\\
\pi_{ab} &=& \gamma^2 \left( \rho^{*} + P^{*} \right) \left( 
v^{a}v^{b} - \frac{v^2}{3} \, h^{ab} \right) \,.
\end{eqnarray}
In the frame $u^a$, the fluid cannot be in equilibrium since 
$q^a\neq 0$: indeed, $q^a = 0 $ implies $ v^c = 0$ and $u^a = 
u^{*a}$.
A perfect fluid is in thermal equilibrium in its comoving frame 
({\em i.e.}, $q^{*a} = 0$), but any other frame moving with 
respect to it ($v^2 > 0$) will experience  a (purely convective) heat flux 
with density  $q^a \neq 0$ given by Eq.~(\ref{ABC}) and there cannot 
be thermal equilibrium.

Before proceeding, let us be clear on the motivations of this work: 
the most interesting applications of the new generalized Tolman-Ehrenfest 
criterion that we present are about conformally invariant systems (the 
cosmic microwave background in cosmology, a 
blackbody gas of Hawking radiation or a massless conformally coupled 
scalar field). It is possible that  useful applications of the new 
criterion will be limited to conformally invariant systems, although 
this is not, by 
all means, established. However, even if this potential 
limitation turns 
out to be real, the generalized Tolman-Ehrenfest criterion of local 
thermal equilibrium presented here is very interesting because 1)~it 
still allows one to discuss interesting (and varied) physics and 
2)~it deepens our 
understanding of thermal physics in relativity. The first point will be 
elaborated in the following sections. As for the second point,  one 
should keep in mind that the 
original Tolman-Ehrenfest criterion, which has not been 
applied widely  to theoretical physics and astrophysics, is
still a valuable contribution to the understanding of thermal physics 
in relativity. The latter is definitely incomplete, on par with 
the understanding of general non-equilibrium thermodynamics. In this 
sense,  generalizing the Tolman-Ehrenfest criterion as done here seems 
valuable for the understanding of local thermal equilibrium in non-static 
spacetimes. 

The rest of this article proceeds as follows. Section~\ref{sec:2} 
discusses two examples showing how to generalize the Tolman-Ehrenfest 
criterion to conformally static spacetimes. Section~\ref{sec:3} derives 
the generalized formula $\tilde{ {\cal T}}={\cal T}/\Omega$ for 
conformally static geometries $\tilde{g}_{ab}=\Omega^2 \, g_{ab}$ in two 
independent ways, while Sec.~\ref{sec:4} discusses an application 
to geometries conformal to the Schwarzschild black hole and 
Sec.~\ref{sec:5} contains a discussion and the 
conclusions.

\section{Examples}
\label{sec:2}
\setcounter{equation}{0}

In this section we examine examples leading to a way of generalizing the 
Tolman-Ehrenfest criterion to conformally static spacetimes 
characterized by the metric $\tilde{g}_{ab} = \Omega^2 \, g_{ab}$, where 
the conformal factor $\Omega \left( x^{\alpha} \right) $ is a regular and 
nowhere-vanishing function of the spacetime coordinates.

\subsection{Example~1: static conformal factor}
\label{subsec:2.1}

The first example is almost trivial but points to the way to proceed in 
more interesting situations. Assume that the metric $\tilde{g}_{ab}$ is 
also static, then there is a timelike Killing\footnote{If the conformal 
factor $\Omega$ is not static, there is only a conformal Killing vector 
$\tilde{k}^a$ in the conformally rescaled spacetime \cite{Wald:1984rg}.} 
vector $\tilde{k}^a$ and, in coordinates adapted to this time symmetry, 
\be
\partial_{t}\, \tilde{g}_{\mu\nu} = 0 \,, \quad \quad  
\tilde{g}_{0i} = 0 \quad (i=1,2,3).
\ee 
The conformal factor is static, $\Omega = \Omega(x^{i})$, hence 
$\partial_t\Omega = 0$ and 
\be
 \partial_{t} \, \tilde{g}_{00} = \partial_t 
\left[ \Omega^2 \left( x^i \right) g_{00}\left( x^i \right) \right] = 0 
\,, 
\quad\quad 
\tilde{g}_{0i} = \Omega^2 g_{0i} = 0 
\ee
and, applying the Tolman-Ehrenfest criterion directly to the static geometry 
$\tilde{g}_{ab}$, one obtains
\be
  \Tilde{ {\cal T} }\sqrt{-\tilde{g}_{00} } = 
\Omega \, \tilde{{\cal T} } \sqrt{-g_{00}} = \mbox{const.}
\ee
Since in the geometry $g_{ab}$ we have $ {\cal T} \sqrt{-g_{00}} = 
$~const., it follows that $  \Omega \, \tilde{{\cal T}} / {\cal T} = 
$~const. One can redefine the time coordinate to absorb the constant  (or 
simply note that $\Omega=1$ must reproduce the identity), obtaining 
\be
    \Tilde{ {\cal T}} = \frac{{\cal T}}{\Omega} \,. \label{newT}
\ee
As we will see in the following, Eq.~(\ref{newT}) relates the temperature 
between conformally related spacetimes also in more physically significant 
situations.

\subsection{Example~2: cosmic microwave background in FLRW universes}
\label{subsec:2.2}

All Friedmann-Lema\^itre-Robertson-Walker (FLRW) universes are conformally 
flat \cite{Wald:1984rg}, hence conformally static. 
Consider, for simplicity, a spatially flat FLRW universe with line element
\begin{eqnarray}
ds^2 &=& -dt^2 + a^2 (t) \left( dx^2 +dy^2 +dz^2\right) \nonumber\\
&&\nonumber\\
&=&  a^2(\eta) \left(-d\eta^2 + dx^2 +dy^2 +dz^2  \right) 
\label{FLRWlineelement}
\end{eqnarray}
in comoving coordinates $\left(t,x,y,z \right)$, or using the conformal 
time  $\eta$ defined by $dt = a(\eta)d\eta$. Consider a radiation 
fluid (the cosmic microwave background) in local thermal 
equilibrium\footnote{Of course, a conformal transformation is just a 
mathematical operation and does not guarantee local thermal equilibrium, 
which must be assumed  and depends on the microphysics (reaction rates 
must be faster than the Hubble rate to  maintain equilibrium 
\cite{KolbTurner}.} in an expanding, spatially flat, FLRW universe.
After decoupling from baryons, the cosmic microwave background evolves as  
a radiation fluid independent of the other fluids present in the 
universe.  It is well-known that, in order to maintain the blackbody 
distribution and thermal equilibrium, the temperature of the cosmic 
microwave background must scale according to ${\cal T} \sim 1/a$ 
\cite{Wald:1984rg}, that is, in accordance with Eq.~(\ref{newT}):
\be
{\cal T}(t) = \frac{ {\cal T}_0}{a(t)} \,,
\ee
where $ {\cal T}_0 = {\cal T} \left( a(t_0) = 1 \right) = {\cal T}(t_0)$ 
is constant and the instant $t_0$ is defined by $a(t_0) = 1$. In 
fact, the 
Planck distribution for the spectral energy density of  a blackbody is
\be
u \left( \nu, {\cal T} \right) = \frac{8\pi h \nu^3}{c^3} 
\, \frac{1}{ \mbox{e}^{\frac{ h\nu }{K_B {\cal T} } } -1} \,,
\ee
where $\nu$ is the photon frequency, ${\cal T}$ the absolute 
temperature, and $h, 
c $, and $K_B$ are the Planck constant, speed of light, and 
Boltzmann constant, respectively.   
Since in a FLRW universe frequencies redshift with the cosmic 
expansion  according to $ \nu 
\sim 1/a $ (equivalently, the proper wavelength scales as $\lambda \, 
a$, where $\lambda $ is the comoving wavelength), it must be $K_B {\cal T}  
\sim 1/a $, or else the Planck distribution would be distorted by the 
cosmic expansion:
\be
   {\cal T} \sim  \frac{1}{a} \sim  \frac{1}{\Omega} 
\ee
where $\Omega = a(\eta)$ is the conformal factor of the FLRW line 
element~(\ref{FLRWlineelement}).

This result can be obtained in another way which highlights formulas 
useful in the following. Assuming the number of photons to be 
conserved (which is true after decoupling), the first law of 
thermodynamics for the radiation fluid reads
\be
   {\cal T}dS = dU +PdV
\ee
where $U$ is the internal energy, $P$ is the radiation pressure, and $V$ 
is the volume. The entropy density is 
\be
    s \equiv \frac{dS}{dV} = \frac{\rho + P}{T}
\ee
where $\rho= dU/dV$ is the energy density, while the entropy is ({\em 
e.g.}, \cite{Carter})
\be
    S = \frac{32\pi^{5}K_B}{45} \left(\frac{K_B {\cal T} }{hc}\right)^3 V 
\,,
\ee
implying that
\be
  \rho + P = \frac{32\pi^5 K^4_B }{45 \left( hc \right)^3} \, {\cal T}^4 
\,.
\ee
For conformal transformations of perfect fluids in FLRW cosmology,  
the pressure and energy density transform as \cite{Faraoni:2004pi}
\be
    \tilde{\rho} = \Omega^{-4}\, \rho \,,\quad\quad  \tilde{P} = 
\Omega^{-4} \, P \,,
\ee
then
\begin{eqnarray}
\tilde{\rho} + \tilde{P} = \Omega^{-4} \left( \rho + P \right) &=& 
\Omega^{_4} \,  \frac{32\pi^5 K^4_B }{45 \left( hc \right)^3 } \, \, {\cal 
T}^4 \nonumber\\
&&\nonumber\\
 = \frac{32\pi^5 K^4_B }{45 \left( hc \right)^3 } \, \, \tilde{{\cal T}}^4 
\,,
\end{eqnarray}
leading again to
\be
    \tilde{ {\cal T} } = \frac{ {\cal T} }{\Omega} 
\ee
where
\be
  {\cal T} = \left[ \frac{45 \left( hc 
\right)^3 }{32\pi^5 K^4_B } \right]^{ 1/4} 
\ee
and where $\rho$ and $P = \rho/3$ are constant for blackbody radiation at 
rest in Minkowski spacetime. 

In other words, one notes that for blackbody radiation 
\begin{eqnarray}
\rho &=&\frac{U}{V}= A {\cal T}^4 \,, \quad \quad A = \frac{8\pi^5 
K_B^4}{15 h^3 c^3} \,,\\
&&\nonumber\\
s &=& \frac{4\rho}{3{\cal T}} \sim {\cal T}^3 \,.
\end{eqnarray}
Then, comparing the expressions of the rescaled energy density 
\begin{eqnarray}
\tilde{\rho} &=& A \tilde{{\cal T}}^4 \,,\\
&&\nonumber\\
\tilde{\rho} &=&  \Omega^{-4} \, \rho \,,
\end{eqnarray}
one obtains Eq.~(\ref{newT}) with $T=\left( \rho/A \right)^{1/4} $. 
These calculations are appropriate to the 
physics at hand: in Minkowski spacetime a 
radiation fluid has $T=$~const. while in FLRW spacetime 
\be
{\cal T} \sqrt{-g_{00}} = \frac{ {\cal T}}{\Omega} \, \Omega 
\sqrt{-g_{00}} = \frac{ {\cal T} \sqrt{ -\tilde{g}_{00}} }{\Omega} =  
\tilde{ {\cal T}} \sqrt{ -\tilde{g}_{00}} \label{Buch-gen}
\ee
implies that $\tilde{ {\cal T}}={\cal T}/\Omega$. The reasoning works 
in coordinates in which $\tilde{g}_{ab}$ is explicitly conformally static, 
that is, comoving frame and conformal time are needed.

\section{Test fluids in conformally static spacetimes}
\label{sec:3}
\setcounter{equation}{0}

We now generalize the Tolman-Ehrenfest criterion for thermal equilibrium 
to the local thermal equilibrium of fluids in conformally static 
spacetimes. That this is possible is suggested by the previous example of 
the cosmic microwave background in FLRW universes. Conformally static 
spacetimes are non-trivial because they can be dynamical (like the FLRW 
geometry), which is a significant deviation from the situation of a static 
fluid at rest in a static spacetime, to which the Tolman-Ehrenfest 
criterion has been confined since its inception 
\cite{Tolman28,Tolman:1930zza,Tolman:1930ona} (only recently a proper 
description of stationary spacetimes has been given 
\cite{Santiago:2018kds,Santiago:2018jeu,Santiago:2019aem}). We provide two 
different derivations of the generalized Tolman-Ehrenfest criterion.

\subsection{Derivation using perfect fluids}

Consider a conformally static metric $\tilde{g}_{ab} = \Omega^2  
\,g_{ab}$, 
where $g_{ab}$ is  static, and use coordinates $\left( t, x^i \right)$  
adapted to the time 
symmetry, in which  $\partial g_{\mu\nu}/\partial t = 0$ and 
$g_{0i} = 0$. The normalization of the four-velocity in the 
conformally rescaled world $  -1= \tilde{u}^c \tilde{u}_c = 
\Omega^{2} \, g_{ab}\tilde{u}^a \tilde{u}^b  $,  in conjunction with   
$g_{ab }u^a u^b = -1$, gives  
\be
\Tilde{u}^c  =  \frac{u^c }{\Omega} \,, \quad \quad \tilde{u}_c = 
\Omega \, u_c \,.
\ee
In the comoving frame of the fluid, assumed to coincide with the 
frame of the static observers, the components of the fluid's 
four-velocity are $u^{\mu} = \left( u^0, 0 ,0,0 \right)$ and 
the conformal image of this frame is the comoving frame of the conformally 
transformed fluid because
\be
    \Tilde{u}^{\mu} = \left(\frac{u^{0}}{\Omega}, 0 ,0,0 \right) \,.
\ee
Denoting with $\Tilde{g}^{(3)}$ the determinant of the spatial 3-metric 
induced by $\Tilde{g}_{ab}$, the three-dimensional volume of a region 
of the rescaled 3-space is 
\be
    \Tilde{V} = \int d^{3}\vec{x} \, \sqrt{\tilde{g}^{(3)}} = \int 
d^{3}\vec{x} \, \sqrt{ \Omega^6 {g^{(3)}}} = \int 
d^{3}\vec{x} \, \Omega^3\sqrt{{g^{(3)}}}
\ee
thus, if $\Omega = \Omega(t)$ then $\Tilde{V} = \Omega^{3}V$, but this is 
not true if $\Omega(x^{\mu})$ depends on the spatial coordinates. However, 
it is always true that for infinitesimal volumes 
$ d\Tilde{V} = \sqrt{\tilde{g}^{(3)} } \, d^3x =
\Omega^3 \, \sqrt{ g^{(3)} } \, d^3x = \Omega^{3}dV $. 
The relations $\Tilde{\rho} = \Omega^{-4}\rho$ and $\Tilde{P} = 
\Omega^{-4}P$ valid for perfect fluids in FLRW spacetimes 
\cite{Faraoni:2004pi} can be 
generalized to test fluids in any conformally static spacetime. In fact, 
equivalent  Lagrangian densities for a perfect fluid are $-\rho$ and $P$ 
\cite{SeligerWhitham,Schutz:1970my,Brown:1992kc, Hawking:1973uf}.  We can 
relate these equivalent actions for a perfect fluid to those of the 
conformally transformed fluid as follows:
\begin{eqnarray}
    J & \equiv & \int d^{4}x \, \sqrt{-g} \, 
\mathcal{L}^\mathrm{(m)}_{(1)} = -\int 
d^{4}x \, \sqrt{-g} \, \rho \nonumber\\
&&\nonumber\\
    & = & -\int d^{4}x \, \sqrt{-\Tilde{g}} \, \Tilde{\rho} 
    =  \int d^{4}x \, \sqrt{-\Tilde{g}} \, 
\Tilde{\mathcal{L}}^\mathrm{(m)}_{(1)} \,,
\end{eqnarray}
where $\Tilde{\rho} = \Omega^{-4}\, \rho$ and $\tilde{g}=\Omega^8 \, g$. 
Similarly, for the equivalent 
perfect fluid Lagrangian,
\begin{eqnarray}
J & \equiv & \int d^{4}x \, \sqrt{-g} \, 
\mathcal{L}^\mathrm{(m)}_{(2)} = \int 
d^{4}x\sqrt{-g} \, P \nonumber\\
    &&\nonumber\\
& = & \int d^{4}x \, \sqrt{-\Tilde{g}} \, \tilde{P} =  
\int d^{4}x \, \sqrt{-\Tilde{g}} \, 
\Tilde{\mathcal{L}}^\mathrm{(m)}_{(2)} \,,
\end{eqnarray}
where $\Tilde{P} = \Omega^{-4} \, P$.
The perfect fluid remains a perfect fluid if we add the information that 
$\Tilde{u}_a = \Omega \, u_a$. In fact,
\begin{eqnarray}
    \Tilde{T}_{ab} &=& \Tilde{\rho} \, \Tilde{u}_a \Tilde{u}_b  
+ \Tilde{P} \tilde{h}_{ab} = \Omega^{-4} \rho \, \Omega \, u_a \Omega 
\, u_b
+ \Omega^{-4} \, P \Omega^2 h_{ab} \nonumber\\
&&\nonumber\\
    &=& \Omega^{-2} \left( \rho u_a u_b + Ph_{ab} \right) 
    = \Omega^{-2} \, T_{ab} \,:
\end{eqnarray}
the conformal transformation does not generate  
dissipative terms in the stress-energy tensor of a test perfect fluid.  
However, if $T_{ab}$ sources the Einstein equations, then 
$\Tilde{g}_{ab}$ is  not a solution of the Einstein equations with 
the same source because these change to
\be 
\Tilde{G}_{ab} =  8\pi(\Tilde{T}_{ab} + T^{(\Omega)}_{ab}) \,,
\ee
where 
\begin{eqnarray}
8\pi T^{(\Omega)}_{ab} &=& -\frac{2}{\Omega} \left( \nabla_a 
\nabla_b \Omega -g_{ab} \Box \Omega \right) \nonumber\\
&&\nonumber\\
&\, & + \frac{1}{\Omega^2} \left( 
4\nabla_a \Omega \nabla_b \Omega-g_{ab} \nabla_c \Omega \nabla^c \Omega 
\right) 
\end{eqnarray}
is generated by $\Omega$ and its first and 
second covariant derivatives.  This fact is immaterial for our discussion, 
in which $T_{ab}$ describes a test fluid and the Tolman-Ehrenfest 
criterion is purely kinematic \cite{Santiago:2019aem}, hence we do not 
worry about the field equations.

Let us proceed with our reasoning. For a conformally static spacetime, 
the proper 3-volume element is $ d\Tilde{V} =
\Omega^3 \, dV \equiv  \Omega^3 \, d\tilde{V}_\mathrm{comoving}$.  For a 
perfect fluid the entropy is constant along the fluid lines, which 
means that there is no entropy generation (because there is no 
dissipation) in the comoving frame, or the entropy remains constant in 
time in the comoving frame and, in this frame,  also the entropy density 
\be
\tilde{s}_\mathrm{comoving} \equiv 
\frac{d\Tilde{S}}{d\Tilde{V}_\mathrm{comoving}} 
= \mbox{const.}
\ee
Then 
\begin{eqnarray}
\tilde{s}_\mathrm{comoving} &=& \frac{d \tilde{S}}{d 
\tilde{V}_\mathrm{comoving}} 
= \frac{d\Tilde{S}}{\Omega^{-3} \, d \tilde{V} } \nonumber\\
&&\nonumber\\
&=& \Omega^3 \, \Tilde{s} = 
\Omega^3 \left(\frac{\Tilde{\rho} + \Tilde{P}}{\Tilde{ {\cal T} }}\right) 
= \mbox{const.}
\end{eqnarray}
Using the fact just proven that $\Tilde{\rho} = \Omega^{-4} \, \rho$, 
$\Tilde{P} = \Omega^{-4} \,P$, we can write
\be
\Omega^{-1}\left(\frac{\rho + P}{\Tilde{ {\cal T}}}\right) = \mbox{const.}
\ee
which implies that
\be
\Tilde{ {\cal T}} = \mbox{const.} \left( \frac{\rho + P}{\Omega} \right)  
= \mbox{const.} \, \frac{ {\cal T}}{\Omega} \left(\frac{\rho + 
P}{ {\cal T}}\right) \,.
\ee
Using now the fact that for the Minkowski space perfect fluid 
$s=\left( \rho+P \right)/ {\cal T}$ is constant, we have $ 
\Tilde{ {\cal T}} = \mbox{const.} \, {\cal T}/ \Omega $. The 
multiplicative constant is determined by the fact that $\Omega=1$ (or more 
generally, $\Omega=$~const.) gives the identity, yielding 
\be
\tilde{ {\cal T}} = \frac{ {\cal T}}{\Omega} \,.\label{Tolmanagain}
\ee

As is well known ({\em e.g.}, \cite{Faraoni:2004pi}), in general the 
stress-energy tensor of the conformally transformed fluid is not 
covariantly conserved but satisfies 
\be
\tilde{\nabla}_b \tilde{T}^{ab} = - \frac{ \tilde{T} \nabla^a 
\Omega}{\Omega^3} 
\ee
and is conserved only for a conformally invariant fluid with 
$T=\tilde{T}=0$ (this is the case for the radiation fluid in FLRW 
universes just considered).

\subsection{Derivation from Eckart's law of heat 
conduction}

In the conformally static geometry $\Tilde{g}_{ab} = \Omega^2 \,  
g_{ab}$, Eckart's law for heat conduction in a dissipative fluid reads 
\cite{Eckart:1940te}
\be    
\Tilde{q}_a = -\Tilde{K} \Tilde{h}_{ab} \left(  
\Tilde{\nabla}^b \Tilde{ {\cal T}} + \Tilde{ {\cal T}} \, \Tilde{a}^b 
\right) \,.
\ee
If the generalized Tolman-Ehrenfest criterion 
$\Tilde{ {\cal T}} = {\cal T}/\Omega$ 
holds in the  conformally rescaled frame, one  should be able to derive 
it directly from Eckart's law~(\ref{Eckart1}) written in this frame, which 
we do 
here. Essentially, we use again the {\em definition} of local thermal 
equilibrium $\tilde{q}^a=0$ and  the temperature 
${\cal T}$ is not required to be time-independent. Indeed, even if 
$ u^c \nabla_c {\cal T}=0$ in the static spacetime, in the rescaled world  
\be
 \tilde{u}^c \tilde{\nabla}_c \tilde{ {\cal T}} = \frac{u^b}{\Omega} \, 
\nabla_b \left( \frac{ {\cal T}}{\Omega} \right) = \frac{ u^b\nabla_b 
{\cal T}}{\Omega^2} -\frac{ {\cal T} u^b\nabla_b \Omega}{\Omega^3} = 
-\frac{ {\cal T} \dot{\Omega}}{\Omega^3} \neq 0 
\ee
unless the conformal factor $\Omega$ is time-independent, which 
would bring us back to the rather trivial example~1 of Sec.~\ref{subsec:2.1}.

Recall that, to derive the Tolman-Ehrenfest law in a static spacetime, 
one uses  the Buchdhal identity \cite{Buchdahl49}  
$a^c  = \nabla^c \ln  \sqrt{-g_{00}} $ (Appendix~\ref{sec:AppendixA}). 
While, in general, it is not true that $\Tilde{a}^c = 
\Tilde{\nabla}^c \ln \sqrt{-\Tilde{g}_{00}} $, the relation
\be
 \tilde{h}_{ab} \, \Tilde{a}^b = 
\tilde{h}_{ab} \, \Tilde{\nabla}^b \ln \sqrt{-\Tilde{g}_{00}} 
\label{eq:relation}
\ee
is valid and is all that is needed. This condition differs from the 
previous one if the four-acceleration has  a component parallel to the 
four-velocity.  Instances in which a four-force 
is parallel to the four-velocity of a particle and causes an {\em 
effective}  
acceleration with the same direction comprise particles with variable mass 
\cite{Mbelek:1998vu,Mbelek:2004ff} (including rockets and solar sails 
\cite{Forward84,Fuzfa:2019djg,Fuzfa:2020dgw}), interacting dark 
energy \cite{Zimdahl:2000zm,Schwarz:2001cf,Zimdahl:1996cg,Zimdahl:1997qe, 
ZimdahlBalakin98a, Zimdahl:1998zq}, timelike geodesic curves 
mapped to the Einstein frame of scalar-tensor or dilaton gravity 
\cite{Dicke:1961gz,Damour:1990tw,Casas:1991ky, 
Garcia-Bellido:1992xlz,Faraoni:2004pi},  
the worldlines of fluid elements in FLRW cosmology as seen by comoving 
observers when the cosmic fluid is not a dust \cite{Faraoni:2020ejhw}, and 
non-affinely parametrized geodesics \cite{EMMacC}.  The reason why a 
four-acceleration is not parallel to the corresponding four-velocity is 
simply because, in these situations, the proper time fails to be an affine 
parameter along the particle trajectory and does not contradict standard 
tenets of special relativity \cite{Faraoni:2020ejhw}.

To prove Eq.~(\ref{eq:relation}), first compute
\begin{eqnarray}
\tilde{a}_a & \equiv & \tilde{u}^b \tilde{\nabla}_b \tilde{u}_a = \frac{ 
u^b}{\Omega} \, \tilde{\nabla}_b \left( \Omega \, u_a \right) = \left( 
\frac{ u^b\nabla_b\Omega}{\Omega} \right) u_a +u^b \tilde{\nabla}_b u_a 
\nonumber\\
&&\nonumber\\
&=& \frac{\dot{\Omega}}{\Omega} \, u_a +u^b \left( \partial_b u_a 
-\tilde{\Gamma}^c_{ab} u_c \right) \nonumber\\
&&\nonumber\\
&=& 
\frac{\dot{\Omega}}{\Omega} \, u_a +u^b \Bigg[ 
\partial_b u_a -\Gamma^c_{ab} u_c \nonumber\\
&&\nonumber\\
&\, &   -\frac{1}{\Omega} \left( 
{\delta^c}_b \partial_a\Omega + 
{\delta^c}_a \partial_b\Omega -
g_{ab} \, \partial^c \Omega \right) \Bigg] u_c \nonumber\\
&&\nonumber\\
&=& \frac{\dot{\Omega}}{\Omega} \, u_a + u^b \nabla_b u_a -
\frac{1}{\Omega} \, u^c u_c \partial_a\Omega \nonumber\\
&&\nonumber\\
&=& \frac{\dot{\Omega}}{\Omega} \, u_a +a_a 
+ \frac{\nabla_a\Omega}{\Omega} \nonumber\\
&&\nonumber\\
& = &  a_a + \frac{1}{\Omega}\left( u_a u_b \nabla^b\Omega  
+ g_{ab} \nabla^b \Omega \right) \nonumber\\
&&\nonumber\\
& \equiv & a_a + h_{ab} \, \frac{\nabla^b\Omega}{\Omega} \,,
\end{eqnarray}
where we used \cite{Wald:1984rg,Faraoni:2004pi}
\be
\tilde{\Gamma}^c_{ab} = \Gamma^c_{ab} +
\frac{1}{\Omega} \left( 
{\delta^c}_b \partial_a\Omega + 
{\delta^c}_a \partial_b\Omega -
g_{ab} \, \partial^c \Omega \right) \,.
\ee

The four-acceleration $\tilde{a}^c$ is still orthogonal to the 
four-velocity $\tilde{u}^c$:
\begin{eqnarray}
\tilde{g}_{ab} \, \tilde{a}^a \, \tilde{u}^b &=& \Omega^{-1} \, 
\tilde{a}_a \, u^a = 
\Omega^{-1} \left( a_a + h_{ab} \, \frac{\nabla^b\Omega}{\Omega} \right) 
u^a \nonumber\\
&&\nonumber\\
&=&   \Omega^{-1} a_b u^b = 0 \,. 
\end{eqnarray}
We now compute 
	%the term $\frac{\dot{\Omega}}{\Omega} \, u_a$ disappears 
	%when projecting it onto the 3-space with ${ \tilde{h}^b}_a$:
\begin{eqnarray}
\tilde{h}_{ab} \, \tilde{a}^b &=& \Omega^2 h_{ab} \left( a^b +h^{bc} \, 
\frac{\nabla_c\Omega}{\Omega} \right) \nonumber\\
&&\nonumber\\
&=& \Omega^2 h_{ab} \left(  \nabla^b \ln \sqrt{-g_{00}} +\nabla^b \ln 
\Omega \right) \nonumber\\
&&\nonumber\\
&=& \Omega^2 \, h_{ab} \nabla^b \ln \left( \Omega \, \sqrt{-g_{00}} 
\right) 
 \nonumber\\
&&\nonumber\\
&=& \Omega^2 h_{ab} \nabla^b \ln  \sqrt{-\tilde{g}_{00}}  \nonumber\\
&&\nonumber\\
&=& \tilde{h}_{ab} \tilde{\nabla}^b \ln \sqrt{-\tilde{g}_{00}} \,,
\end{eqnarray} 
which completes the proof\footnote{Contrary to the proof of the 
analogous relation for static spacetimes (Appendix~\ref{sec:AppendixA}), 
the Killing equation has not been used. Indeed, the conformally rescaled 
world, in general, has no timelike Killing vector, but only a conformal 
Killing 
vector \cite{Wald:1984rg}.} of Eq.~(\ref{eq:relation}).  One then has 
\begin{eqnarray}
\tilde{q}_a &=& -\tilde{ {\cal K}} \, \tilde{h}_{ab} \left( 
\tilde{\nabla}^b \tilde{ {\cal T}} 
+\tilde{ {\cal T}} \, \tilde{a}^b \right) \nonumber\\
&&\nonumber\\
&=& -\tilde{ {\cal K}} \, \tilde{h}_{ab} \tilde{ {\cal T}}  
\left( \tilde{\nabla}^b \ln \tilde{ {\cal T}} 
+ \tilde{\nabla}^b \ln \sqrt{ -\tilde{g}_{00} }  \right) \nonumber\\
&&\nonumber\\
&=& -\tilde{ {\cal K}} \, \tilde{h}_{ab} \tilde{ {\cal T}}  
\tilde{\nabla}^b \ln \left( \tilde{ {\cal T}} \, \sqrt{ -\tilde{g}_{00}}  
\right) 
\end{eqnarray}  
and thermal equilibrium $\tilde{q}_a=0 $ implies that 
$ \tilde{\nabla}^b \ln \left( \tilde{ {\cal T}} \, \sqrt{ -\tilde{g}_{00}}  
\right) $ is parallel to $\tilde{u}^b$. Then $ \tilde{ {\cal T}} \, \sqrt{ 
-\tilde{g}_{00}} $ must depend only on time,
\be
\tilde{ {\cal T}} \, \sqrt{ -\tilde{g}_{00} } =f(t) 
\ee
where $f(t)$ is an integration function, or
\be
\tilde{ {\cal T}} = \frac{ f(t) }{\Omega \sqrt{-g_{00}} } = 
\frac{ f(t) {\cal T} }{\Omega \left( {\cal T} \,\sqrt{-g_{00} } \right)} =
\mbox{const.} \, \frac{ f(t) {\cal T} }{\Omega } \,.
\ee
The product $ \mbox{const.} \times f(t)$ is fixed by the fact that, if 
$\Omega \equiv 1$, the conformal transformation must reduce to the 
identity with $ \tilde{ {\cal T}} ={\cal T}$ and $ \mbox{const.} \times 
f(t)=1 $. We are left with Eq.~(\ref{Tolmanagain}) again.

\section{Conformally Schwarzschild geometries}
\label{sec:4}
\setcounter{equation}{0}

It is interesting to compare the generalized Tolman-Ehrenfest 
criterion~(\ref{newT}) with spacetimes designed intentionally to be 
conformal to the (exterior) Schwarzschild black hole geometry
\be
ds_{(0)}^2 = -\left( 1-\frac{2m}{r} \right) dt^2 + \frac{dr^2}{1-2m/r} 
+r^2 d\Omega_{(2)}^2 \,,
\ee
where $d\Omega_{(2)}^2\equiv d\vartheta^2 +\sin^2 \vartheta \, d\varphi^2$ 
is the line element on the unit two-sphere and the parameter $m$ is the 
(constant) black hole mass. 

The first such spacetime described here is the Sultana-Dyer solution of 
the Einstein equations, which is a Petrov type D, time-dependent, and 
spherically symmetric spacetime sourced by two non-interacting fluids, a 
null dust and an ordinary (timelike) dust \cite{Sultana:2005tp}. It is 
interpreted as a describing a black hole embedded in a spatially flat FLRW 
universe.

Since we need a test fluid at rest in the static seed spacetime to apply 
the criterion~(\ref{newT}), we consider the region around the 
Schwarzschild event horizon, in which Hawking radiation creates a static 
blackbody radiation fluid at the Hawking temperature ${\cal 
T}=\frac{1}{8\pi m}$ (in geometrized units). The Tolman-Ehrenfest  
criterion clearly 
fails at horizons since, for the Schwarzschild black hole it would give 
${\cal T}=\frac{{\cal T}_0 }{1-2m/r}$, which diverges as $r\to 2m^{+}$. 
However, the cause is not that the criterion is inherently bad but  
it is restricted to static coordinates, and the latter fail at the 
Schwarschild event horizon. Hawking radiation is a quantum  
phenomenon and the proper calculation of the Hawking temperature requires 
quantum field theory in curved space, including a careful consieration of 
the vacuum state. Once this is done and the temperature appearing in the 
Tolman-Ehrenfest criterion is cured producing the Hawking result ${\cal 
T}=\frac{1}{8\pi m}$, one can consider the Schwarzschild geometry as a 
seed for constructing the Sultana-Dyer spacetime by a conformal 
transformation. The Sultana-Dyer line element is \cite{Sultana:2005tp}  
\begin{eqnarray}
ds^2 &=& a^2 \left(\eta, r \right) \left[ -\left( 1-\frac{2m}{r} \right) 
d\eta^2 + \frac{dr^2}{1-2m/r} +r^2 d\Omega_{(2)}^2 \right] \nonumber\\
&&\nonumber\\
&=&  a^2 \left(\eta, r \right) ds_{(0)}^2 \,,
\end{eqnarray}
where
\be
a \left(\eta, r \right) = \left( \eta +2m \ln \Bigg| \frac{r}{2m} -1 
\Bigg| \right)^2 \,.
\ee
If $m=0$ the line element reduces to the FLRW one written in conformal 
time. The coordinate change
\be
\tau \left(\eta, r \right) = \eta 
+  2m \ln \Bigg| \frac{r}{2m} -1 \Bigg| 
\ee 
turns the line element into the original Sultana-Dyer form 
\cite{Sultana:2005tp} 
\be 
ds^2 = a^2 (\tau ) \left[ 
  - d\tau^2 + dr^2 +r^2 d\Omega_{(2)}^2 -\frac{2m}{r} \left( d\tau + dr 
  \right)^2 \right] \,, 
\ee 
with $a(\tau)=\tau^2$ \cite{Sultana:2005tp}. The Tolman criterion applied 
to the Sultana-Dyer 
geometry yields the temperature 
\be {\cal T}=\frac{ {\cal T}_0}{a \sqrt{ 1-2m/r} } 
\ee 
which, as usual, diverges at the event horizon where the static 
coordinates fail, and needs to be regularized. This has been done 
by Saida, Harada, and Maeda \cite{Saida:2007ru}, who studied the Hawking 
radiation of a massless, conformally coupled scalar field $\phi$ in this 
geometry and computed the renormalized stress-energy tensor $\langle 
T_{ab} \left[ \phi \right]  \rangle$ taking into account the conformal 
anomaly and particle creation.  The calculation, analogous to Hawking's 
calculation in the fixed Schwarschild geometry with constant mass $m$ 
(that is, neglecting backreaction), is feasible only in an adiabatic 
approximation in which the black hole mass evolves very slowly, which is 
necessary to guarantee thermal equilibrium. (This condition is analogous 
to the condition that reaction rates exceed the Hubble expansion rate in a 
FLRW universe to  maintain local thermal equilibrium.) The generalized 
Tolman-Ehrenfest criterion~(\ref{newT}) then predicts that the temperature 
of the 
Sultana-Dyer black hole is ${\cal T}={\cal T}_0/\Omega={\cal T}_0/a$, 
where ${\cal T}_0$ is the Hawking temperature. The calculation of 
\cite{Saida:2007ru} produces the result
\be
{\cal T}= \frac{1}{8\pi m a} + \, ... \label{SHM}
\ee
where the corrections omitted are negligible in the adiabatic 
approximation of a slowly evolving black hole \cite{Saida:2007ru}. 

An independent calculation using the method of chiral anomaly confirms the 
temperature~(\ref{SHM}) of the Sultana-Dyer black hole 
\cite{Majhi:2014hpa,Bhattacharya:2016kbm}, which is supported also by 
previous heuristic dimensional reasoning \cite{Faraoni:2007gq}. The 
generalized Tolman-Ehrenfest criterion makes a definite prediction about 
the temperature of cosmological black holes conformal to Schwarzschild, 
and the conformal transformation is a popular technique to generate exact 
solutons of general relativity \cite{Faraoni:2015ula} and of alternative 
theories of gravity \cite{Faraoni:2021nhi}.

\section{Conclusions}
\label{sec:5}
\setcounter{equation}{0}

The applicability of the Tolman-Ehrenfest criterion~(\ref{Tolman}) for the 
thermal equilibrium of a fluid is quite restricted. It requires a static 
spacetime and a fluid at rest with respect to the static observers of the 
latter, who have four-velocity parallel to the timelike Killing vector.  
Extending the Tolman-Ehrenfest criterion to more general geometries is, 
therefore, not an insignificant task. Here we have presented its 
generalization to conformally static spacetimes with metric 
$\tilde{g}_{ab}=\Omega^2 \, g_{ab}$, where $g_{ab}$ is static and the test 
fluid of temperature ${\cal T}$ is at rest in the frame associated with 
the static observers of $g_{ab}$. Then, assuming local thermal 
equilibrium, the generalization of the Tolman-Ehrenfest criterion  
is $ \tilde{ {\cal T}}= {\cal T}/\Omega$. The 
most obvious application of this generalized criterion is to the cosmic 
microwave background in FLRW universes, which reproduces the well known 
scaling of its blackbody temperature $ {\cal T} \sim 1/a$.

The temperature scaling $\tilde{ {\cal T}}={\cal T}/\Omega$ found is 
compatible with Dicke's heuristic argument on the scaling of physical 
quantities under conformal  transformations \cite{Dicke:1961gz}  
(cf. Ref.~\cite{Faraoni:2007gq}) and is confirmed by precise calculations 
\cite{Majhi:2014hpa,Bhattacharya:2016kbm} in the particular case of the 
Sultana-Dyer black hole, as discussed in the previous section.  With this 
argument, physical 
quantities themselves do not carry physical meaning, which is instead 
attributed to the ratios of physical quantities to their units, the only 
outcome of measurements. Usually the units are taken to be constant in 
spacetime, but a conformal rescaling amounts to a rescaling of physical 
units that depends on the spacetime point: lengths and times scale as 
$\Omega$, masses scale as $1/\Omega$, and derived quantities scale 
accordingly to their dimensions \cite{Dicke:1961gz}. Then, since $K_B 
{\cal T}$ is an energy, or a mass, and $K_B$ remains constant, ${\cal T}$ 
should scale as $1/\Omega$, which is what we found. Dicke's argument, 
however, is rather heuristic and is known to become imprecise in the  
conformal transformation from Jordan to Einstein frame in scalar-tensor 
gravity. One must be precise in the discussion 
of what kind of fluid is considered, according to which observers, the 
definition of local thermal equilibrium, and the vanishing of $q^a$ and 
$\tilde{q}^a$.  It is interesting, however, that our finding agrees with 
Dicke's heuristic reasoning.

Already the generalization of the Tolman-Ehrenfest criterion to 
stationary, but non-static, spacetimes requires much care 
\cite{Santiago:2018kds,Santiago:2018jeu,Santiago:2019aem}. The extension 
of our generalization to conformally stationary spacetimes is problematic 
because, under conformal transformations, the non-unique timelike Killing 
vector of the stationary spacetime $g_{ab}$ does not map into another 
timelike Killing vector of $\tilde{g}_{ab}$, but only into a conformal 
Killing vector \cite{Wald:1984rg}. In any case, the problems encountered 
in stationary but non-static spacetimes 
\cite{Santiago:2018kds,Santiago:2018jeu,Santiago:2019aem} are not going to 
be  cured in conformally stationary ones. 

The most interesting applications of the generalized 
Tolman-Ehrenfest criterion~(\ref{newT}) uncovered here (the cosmic 
microwave background in FLRW universes and the Hawking temperature of the 
Sultana-Dyer black hole) are about conformally invariant systems (a 
blackbody gas of Hawking radiation or a massless conformally coupled 
scalar field, which is conformally invariant \cite{Wald:1984rg}). We 
suspect that the most useful applications of this criterion will involve 
conformally invariant systems, but other applications are not excluded at 
this stage. Even with this potential restriction, however, it appears that 
interesting physics can be tackled with the new criterion~(\ref{newT}). 
Indeed, the phenomena discussed here are already quite varied, ranging 
from cosmology to time-dependent black holes. In any case, even the 
original Tolman-Ehrenfest criterion of thermal equilibrium now reported in 
textbooks \cite{MTW} has not found widespread applications to theoretical 
physics and astrophysics, but it has intellectual value in itself as a 
contribution to the understanding of thermal physics in relativity, which 
is still fairly incomplete (as is the understanding of non-equilibrium 
thermodynamics in general), and its generalization to non-static 
situations appears to be valuable.

Finally, in our derivation we used the 
fact that the fluid is a {\em test} fluid. Although the original 
Tolman-Ehrenfest temperature gradient is a kinematic effect, relating 
solutions of the Einstein equations (or of the field equations of 
alternative theories of gravity) through conformal transformations spoils 
the reasoning of Sec.~\ref{sec:3} (an exception is the radiation fluid 
which, due to its conformal invariance and the fact that photons are 
massless, is conserved after the conformal transformation). Further 
generalization of the Tolman-Ehrenfest criterion beyond conformally static 
spacetimes and test fluids seems difficult to achieve.

\begin{acknowledgments}

We thank two referees for useful comments. This work is supported, in 
part, by the Natural Sciences \& Engineering Research Council of Canada 
(grant 2016-03803 to V.F.).

\end{acknowledgments}

\appendix
\section{Derivation of the Tolman-Ehrenfest criterion from Eckart's 
law~(\ref{Eckart1})}
\label{sec:AppendixA}
\renewcommand{\theequation}{A.\arabic{equation}}
\setcounter{equation}{0}

Here we derive the Tolman-Ehrenfest criterion from Eckart's generalization 
of the 
Fourier law for heat conduction in imperfect fluids \cite{Eckart:1940te}, 
using modern notation. We follow Ref.~\cite{Santiago:2019aem} 
step-by-step.
 
Consider  a static test fluid at rest in a static 
spacetime and let $g_{\mu\nu}$ be the metric components in coordinates 
adapted to the time symmetry. The timelike Killing vector $k^a$ satisfies 
the Killing equation
\be
 \nabla_{(a } k_{b)} = \frac{1}{2} \left( \nabla_{a}k_{b} + 
\nabla_{b}k_{a} \right) = 0 \label{Killingequation}
\ee
and has  components   
$k^\mu = \left( 1,0,0,0 \right)$ in these coordinates, while $g_{00} = 
k_c  
k^c $. 

The first step consists of a relation, due to Buchdahl \cite{Buchdahl49}, 
between 
the four-acceleration of a test particle at rest with respect to the 
static observers and $g_{00}$, 
\be
a_c \equiv \dot{u}_c \equiv u^b \nabla_b u_c = \nabla_c \ln \left( 
\sqrt{-g_{00}} 
\right) \,.
\ee
To prove this relation, note that a fluid element has normalized 
four-velocity
\be
    u^{a} = \frac{k^a}{\sqrt{-k^b k_b}}
\ee
and four-acceleration 
\begin{align}
    \frac{du_b}{d\tau} & \equiv u^{c}\nabla_{c}u_{b} 
     = u^{c}\nabla_{c}\left(\frac{k_{b}}{\sqrt{-k^{d}k_{d}}}\right) 
    \nonumber\\
	&\nonumber\\
    & = u^{c}\left[\frac{\nabla_{c}k_{b}}{\sqrt{-k^{d}k_{d}}} + 
    k_b \left(\frac{-1}{2}\frac{1}{(-k^{d}k_{d})^{3/2}}\right) 
   \nabla_{c}(-k^{d}k_{d})\right] \nonumber\\
   &\nonumber\\
 & = u^{c}\left[\frac{\nabla_c  k_b  }{\sqrt{-k^ d k_d }} - 
 \frac{k_b \nabla_c \left( 
  -k^d k_d \right) }{2 \left( -k^d k_d \right)^{3/2}}\right] 
\end{align}
(where $\tau$ is the proper time along the fluid lines).  
The second term in the last line vanishes since 
$k^a \nabla_a\left( k_b k^b \right) = 0$ because 
\begin{eqnarray*}
    k^a \nabla_a \left( k_b k^b \right)
    & = & k^a \nabla_a \left(g_{bc}k^b k^c \right) \\
	&& \\
    & = & k^a g_{bc} \left( k^c \nabla_a k^b + k^b \nabla_a k^c \right) \\
	&& \\
    & = & k^a k^c \left( \nabla_a k_c \right) + 
        k^a k^c \left( \nabla_a k_c \right) \\
	&& \\
    & = & 2k^a k^c \nabla_a k_c \\
	&& \\
    & = & 2k^a k^c \nabla_{(a}k_{c)} = 0 
\end{eqnarray*}
where, in the second to last step, we used the fact that since 
$k^{a}k^{b}$ is symmetric only the symmetric part of 
$\nabla_{a}k_{b}$ contributes, while the last step follows from 
the Killing equation. Then the fluid's four-acceleration is 
\begin{eqnarray}
a_{b} &=& \frac{ u^c \nabla_c k_b }{ \sqrt{-k^d k_d }}
    = \frac{k^c}{\sqrt{-k^d k_d }} \, \frac{\nabla_c  
   k_b}{ \sqrt{-k^d k_d }} \nonumber\\
&&\nonumber\\
&=& \frac{k^c \nabla_c k_b }{-k^d k_d }
    = \frac{-k^c \nabla_b k_c }{-k^d k_d } \,,
\end{eqnarray}
where we used again the Killing equation~(\ref{Killingequation}). Since $  
\nabla_{b}(k^{c}k_{c}) = 2k^{c}\nabla_{b}k_{c}$, 
\be
    k^{c}\nabla_{b}k_{c} = \frac{1}{2} \, \nabla_{b} \left( k^c k_c 
\right)
\ee
and the above identity yields 
\be
    a_{b} = \frac{\nabla_c \left( -k^d k_d \right)}{-2k^d k_d }
    = \frac{ \nabla_{b}\ln \left( -k^d k_d \right)}{2} 
    = \nabla_{b}\ln \left( \sqrt{-k^d k_d } \, \right) \,.
\ee
In the adapted coordinates $k^{a}k_{a} = g_{00}$, hence
\be
    a_{b} = \nabla_{b}\ln\sqrt{-g_{00}}  \,.
\ee
Eckart's generalization of the Fourier law for heat conduction in 
dissipative fluids is then used to complete the derivation. The 
Tolman-Ehrenfest  
criterion refers to {\em perfect} fluids, described by  the 
simpler stress-energy tensor
\be
    T_{ab} = {\rho}u_{a}u_{b} + Ph_{ab} \,,\label{perfectfluid}
\ee
but in Eckart's first-order thermodynamics an imperfect fluid at rest 
coincides with that of a perfect fluid because the imperfect fluid 
dissipative quantities are assumed to be linear in the gradient of the 
four-velocity 
(cf. Eqs.~(\ref{Eckart1})-(\ref{Eckart3}) \cite{Eckart:1940te}. 

For a fluid at rest in a static spacetime, $\Theta$ and $\sigma_{ab}$ 
vanish and $q^a = 0$ in thermal equilibrium, hence the stress-energy 
tensor~(\ref{imperfectfluid})  takes the perfect fluid 
form~(\ref{perfectfluid}).  The temperature of such a fluid is 
time-independent,
\be
\frac{ d {\cal T} }{ d\tau} \equiv u^a \nabla_a {\cal T} = 0 
\ee
then,
\be
  h_{ab}\nabla^b {\cal T} \equiv  
\left( u_a u_b  + g_{ab} \right) \nabla^b {\cal T} =  
\nabla_a {\cal T} \,.
\ee\\
By definition there is no heat flow in thermal equilibrium, $q_{a} = 0$, 
and Eckart's law~(\ref{Eckart1}) gives
\begin{eqnarray*}
\nabla_a {\cal T} + {\cal T}\, \nabla_a \left( \ln{\sqrt{-g_{00}}} 
\, \right) 
= 0 \,,\\
\\
\nabla_a \ln  {\cal T} +\nabla_a \ln {\sqrt{-g_{00}} }  = \mbox{const.}\,, 
\end{eqnarray*}
and finally
\be
{\cal  T} \sqrt{-g_{00}} = \mbox{const.}
\ee

\end{document}